\documentclass[aps,pra,showpacs,notitlepage]{revtex4-1}

\usepackage{graphicx}

\begin{document}

\title{Quantum causality relations and the emergence\\
of reality from coherent superpositions}

\author{Holger F. Hofmann}
\email{hofmann@hiroshima-u.ac.jp}
\affiliation{
Graduate School of Advanced Sciences of Matter, Hiroshima University,
Kagamiyama 1-3-1, Higashi Hiroshima 739-8530, Japan
}

\begin{abstract}
The Hilbert space formalism describes causality as a statistical relation between initial experimental conditions and final measurement outcomes, expressed by the inner products of state vectors representing these conditions. This representation of causality is in fundamental conflict with the classical notion that causality should be expressed in terms of the continuity of intermediate realities. Quantum mechanics essentially replaces this continuity of reality with phase sensitive superpositions, all of which need to interfere in order to produce the correct conditional probabilities for the observable input-output relations. In this paper, I investigate the relation between the classical notion of reality and quantum superpositions by identifying the conditions under which the intermediate states can have real external effects, as expressed by measurement operators inserted into the inner product. It is shown that classical reality emerges at the macroscopic level, where the relevant limit of the measurement resolution is given by the variance of the action around the classical solution. It is thus possible to demonstrate that the classical notion of objective reality emerges only at the macroscopic level, where observations are limited to low resolutions by a lack of sufficiently strong intermediate interactions. This result indicates that causality is more fundamental to physics than the notion of an objective reality, which means that the apparent contradictions between quantum physics and classical physics may be resolved by carefully distinguishing between observable causality and unobservable sequences of hypothetical realities ``out there''.
\end{abstract}

\maketitle

\section{Introduction}

Since the earliest days of quantum mechanics, the representation of quantum states as a superposition of possible measurement outcomes has caused much confusion and controversy. On the one hand side, it is natural to assume that measurement outcomes should somehow correspond to ``elements of reality,'' independent of whether they are actually observed or not. On the other hand side, quantum superpositions do not allow any intermediate measurements of the Hilbert space components that make up the quantum state, lest the measurement destroy the phase relations that are needed to fully define the distribution of future measurement outcomes. It may be tempting to treat the quantum state as a description of reality, but it should be recognized that it is merely a representation of the initial conditions that allows us to calculate the probabilities of future events caused by these initial conditions. The viewpoint I will take in the following is that the quantum state is a representation of statistics, where Hilbert space components represent potential measurement outcomes. Reality should be defined in terms of such measurement outcomes,  and the only justification of superpositions is that their phase relations determine
the probabilities of such measurement outcomes. The problem associated with this understanding of quantum theory is that individual quantum states describe a potentially infinite number of different and mutually incompatible measurement contexts. Instead of limiting itself to the characterization of quantum states, a scientific discussion of the physical meaning of quantum theory should therefore focus on the reproducible causality relations between initial conditions determined by the external manipulation of the system and final conditions observed as a result of a measurement performed on the same system. A number of approaches investigating such causality relations between input states and output measurements have been proposed within the framework of quantum information theory, mostly based on an operational approach to measurement theory \cite{Zei99,Bru99,Fuc03,Cav07,Goy08,Lee11,Lei13,Hof20}. Unfortunately, these approaches tend to treat quantum systems as black boxes with unspecified physical properties, thereby obscuring the relation between experimentally observed statistical correlations and the underlying physics of reproducible relations between cause and effect associated with the emergence of classical equations of motion. 

It is a natural expectation that physics should tell us something about the laws that govern the dynamics of a system, and these laws should be sufficiently objective so that they can be formulated without any reference to the measurement context. This is precisely the purpose of the Hilbert space formalism: it provides an objective description of causality that can be applied to any combinations of state preparation and measurement. The main difference to the corresponding classical description of causality in phase space is the use of quantum superpositions which prevents us from attributing reality to intermediate observations. This seems to be at odds with our macroscopic experience, since we can normally watch objects as they move. In order to understand the relation between classical theories and quantum mechanics properly, it would seem to be important to understand how quantum mechanics reconciles the experience of objective causality as an undisturbed sequence of observations with the impossibility of precise intermediate measurements between an initial condition and a final outcome. 

The problem of intermediate measurements and the associated resolution-disturbance trade-off have been widely discussed in the literature, and recent advances in quantum information technologies have led to a number of successful experimental demonstrations of some of the stranger aspects of quantum measurements \cite{Res04,Jor06,Lun09,Yok09,Gog11,Suz12,Den14,Oka16,Min19}. As a result, we now have a wide range of theoretical and experimental tools that may allow us to address fundamental questions regarding the objective physics described by the quantum formalism. Of particular interest is the question of whether we can observe physical properties of a quantum system without disturbing the time evolution of the system between its preparation and a final measurement. An important breakthrough in that direction has been achieved using weak measurements \cite{Aha88}, which has resulted in a direct observation of quantum coherences between an initial preparation and a final measurement \cite{Wis02,Hof10a,Hof10b,Hos10,Bed10,Lun11,Hof11,Lun12,Hof12a,Mor13,Das14,Dre14a}. In these weak measurement, quantum coherences are observed as an average shift of the meter system, lending some credibility to the idea that we might be looking at an intermediate reality that manifests itself as a well-defined external effect. The problem with the weak measurement limit is the low resolution of each individual measurement outcome, which makes it highly problematic to associate the average result with a physical property of the individual system \cite{Hof12b,Bed13,Mac14,Moch14,Ips14,Coh18,Mat19}. It is therefore of great interest to consider alternative approaches that can provide meaningful insights into the causal origin of intermediate observations while keeping the disturbance of the measurement interaction at a negligibly low level. 

In this paper, I consider the possibility of sidestepping the very tight relation between resolution and disturbance for the initial quantum state by considering only the causality relation between an input $\mid a \rangle$ and a final result $\mid b \rangle$. It is then possible to identify a quantitative condition for disturbance free measurements that depends on the specific relation between the initial state and the measurement result described by the inner product of the two. For sufficiently large quantum systems, this relation can be characterized by the quantum phases of intermediate states which can be expressed as an action $S(a,m,b)$. It can then be shown that the classical notion of an intermediate reality emerges from the contribution $m$ that minimizes this action. The analysis below thus demonstrates that our classical notion of reality is an emergent feature of quantum measurements performed with a sufficiently low resolution. Importantly, the action $S(a,m,b)$ defines the intermediate measurement resolution with respect to the fundamental constant $\hbar$, explaining the classical limit as a natural approximation of the underlying quantum formalism. The results presented in the following demonstrate that the classical notion of reality is based on an emergent phenomenon which is not fundamental to physics. This means that the notion of objective reality can be discarded as a redundant feature, even where the classical explanation of physical phenomena is concerned. Instead, the notion of observable reality can be rooted in fundamental causality relations that are sufficiently accurate to give an objective meaning to subjective experience.

\section{Causality of input-output relations}

It seems that the textbook explanations of quantum mechanics have resulted in the widely held misconception that the time evolution of a quantum state described by Schr\"odinger's equation or by path integrals solves the problem of causality in quantum physics. However, it needs to be recognized that the hypothetical intermediate states of such theories are completely unobservable and have absolutely no empirical validity. Instead, they merely serve as efficient book keeping tools in a theory based on the assumption that a final measurement of the state is possible. It would therefore be a fallacy to interpret the continuous time evolution of a state as a description of the actual dynamics of an individual quantum system. The supposed agreement between these theories and experiment is entirely based on the observation of input-output relations between initial quantum states and final measurements. The only valid representation of empirical causality in quantum mechanics is therefore given by the probability of observing a measurement outcome $b$ after implementing initial conditions $a$,
\begin{equation}
\label{eq:inout}
P(b|a) = |\langle b \mid a \rangle|^2.
\end{equation}
The position I take in the following is that measurement probabilities are absolutely necessary to attach any physical meaning to the mathematical formalism. This position rejects the belief that quantum physics can have a physical meaning independent of Born's introduction of the statistical interpretation embodied by Eq.(\ref{eq:inout}). Specifically, I do not think that it is possible to identify any physical meaning of the quantum formalism without at least an implicit reference to a measurement process with an unpredictable outcome. It is therefore an essential requirement of empirical science that the complete mathematical formalism must be justified in terms of the statistics predicted by Eq.(\ref{eq:inout}), and not by untestable speculations about the ontological status of mathematical expressions. 

If these statements appear a bit harsh to the reader, I would like to point out that this viewpoint is the implicit assumption behind all measurement based interpretations of quantum mechanics, which includes most of the interpretations based on quantum information theory \cite{Zei99,Bru99,Fuc03,Cav07,Goy08,Lee11,Lei13}. The purpose of the present paper is to bridge the gap between theories that recognize only measurement outcomes as real and theories that completely neglect the role of measurement. The problem with the measurement based quantum information approach is that it treats the quantum system as a black box that magically produces the measurement outcomes $b$ from initial conditions $a$. In terms of information, the only role of the Hilbert space vectors $\mid a \rangle$ and $\mid b \rangle$ is to represent inputs and outputs, without reference to the physical properties of the quantum system that might provide additional details about this relation. The question then should be whether this approach can tell us anything at all about the actual physics going on inside this black box. 

To answer this question affirmatively, it is necessary to acknowledge that the inner product in Eq. (\ref{eq:inout}) is not just a statement about an input-output relation, but expresses a fundamental characteristic of the quantum system described by the Hilbert space formalism. It should be noted that the time evolution of the quantum state is completely contained in the formulation given above, since the invasive action of state preparation is usually finished at a time $t_a$ and the equally invasive measurement occurs at a later time $t_b$. We can therefore use the unitary time evolution of a closed system to express the relation between state vectors and the time evolution of the system. In Eq.(\ref{eq:inout}), time has been eliminated by including it in the definition of the states. If the states $\mid a \rangle$ and $\mid b \rangle$ are defined with reference to a standard time $t_m$, their time evolution can be expressed as
\begin{eqnarray}
\label{eq:time}
\mid a(t) \rangle &=& \hat{U}(t-t_m) \mid a \rangle
\nonumber \\
\mid b(t) \rangle &=& \hat{U}(t-t_m) \mid b \rangle.
\end{eqnarray}
The conventional formulation of unitary time evolution can then be recovered by chosing the preparation time $t_a$ for $a$ and the measurement time $t_b$ for $b$,
\begin{equation}
P(b|a) = |\langle b(t_b) \mid \hat{U}(t_b-t_a)\mid a(t_a) \rangle|^2.
\end{equation}
This reformulation shows that the complete internal time evolution of the system is already accounted for by the relation of the Hilbert space vectors $\mid a \rangle$ and $\mid b \rangle$ to each other. It is therefore possible to explain causality without referring to a continuous time evolution of the quantum states. From a quantum information perpective, the time evolution has no physical meaning since it has no effect on the output data produced in the experiment. In the context of weak measurements, it is particularly important that the measurement result $\mid b \rangle$ propagates backwards in time in the same manner that the initial state $\mid a \rangle$ propagates forward in time \cite{Aha88}. Here, I propose to avoid the problems associated with continuous quantum state evolutions by focusing on the causality of observable effects at specific times. This makes it possible to replace hypothetical assumptions about unobserved ``paths'' with the more specific question of how the Hilbert space formalism represents the relations between the physics observed at time $t_m$ and the physics observed at time $t_b$. 

\section{Relation between incompatible measurements}

In terms of the conventional formulation of quantum dynamics, the measurement of $\mid b \rangle$ at time $t_b$ is incompatible with a measurement of $\mid m \rangle$ at time $t_m$. It is therefore impossible to observe the dynamics of a system moving from $a$ through $m$ to $b$. In terms of the Hilbert space algebra, the time evolution of a quantum system is represented by an infinite number of non-commuting operators, making it fundamentally impossible to observe the dynamics of the system. The only experimental evidence we have originates from a comparison of separate measurements, where one set of measurements is performed at time $t_b$ and another is performed at time $t_m$. In the path integral formalism, the final probability is calculated by interfering an infinite number of ``paths'' formed by hypothetical sequences of states associated with the non-commuting operators at different times. As shown in \cite{Hof12a}, this is merely a redundant combination of infinitesimal transformations that are more efficiently expressed by unitary operators. The only physically meaningful question is how the Hilbert space formalism relates the outcomes of one measurement to the outcomes of another measurement when non-commutativity makes it impossible to perform the two measurements jointly. 

To solve this problem, we will now concentrate on the relation of the measurement probabilities given in Eq.(\ref{eq:inout}) with the possibility of an alternative measurement represented by an orthogonal basis set $\{\mid m \rangle\}$ at an intermediate time $t_m$. It is of course well known that this relation is mathematically expressed by the expansion of the Hilbert space inner product in the $m$-basis,
\begin{equation}
\label{eq:expand}
\langle b \mid a \rangle=\sum_m \langle b \mid m \rangle \langle m \mid a \rangle.
\end{equation}
This is the origin of the widespread notion that quantum states represent superpositions of ``realities,'' which seems to be an over interpretation of the basis vectors $\mid m \rangle$. Proceeding more carefully, we can merely say that the causality relation between $a$ and $b$ in Eq.(\ref{eq:inout}) can be related to elements of the causality relations between $a$ and $m$ and between $m$ and $b$, as represented by their respective inner products. However, only the absolute squares of these inner products have a proper physical meaning, which is problematic because the summation over several inner products is highly sensitive to the quantum phases that occur in the expansion. To fully understand this problematic nature of the quantum formalism, it is important to consider the relation between Eq.(\ref{eq:expand}) and the traditional attempts to visualize quantum physics in terms of interferences between different ``realities.'' Clearly, the causality relation between $a$ and $b$ depends on the phase differences between the different values of $m$ associated with "paths" between $a$ and $b$. However, the classical limit of causality suggests that the combination of initial condition $a$ and final condition $b$ should select a specific value of $m$ as a function of $(a,b)$. It should therefore be sufficiently obvious that the causality relating $a$ to $b$ is not determined by ``paths'' through intermediate values of $m$. Nevertheless, unitary transformations are just as deterministic as classical laws of motion. Indeed, the classical laws of motion must somehow emerge from quantum physics, even though the classical notion of intermediate realities cannot be salvaged. Here, I propose to solve this problem by investigating the precise limitations on intermediate measurements introduced by the role of quantum superpositions in defining the input-output relations in Eq.(\ref{eq:inout}).

In the context of Hilbert space inner products, the relation between classical causality and quantum phases can be identified using the action of unitary transformations \cite{Hof11,Hof14,Hof16,Hib18}. As pointed out in \cite{Hof11}, the action $S(a,m,b)$ can be defined as the action of the unitary transformation with eigenstates $\mid m \rangle$ that maximizes the inner product given by 
\begin{equation}
\label{eq:unitary}
\langle b \mid \hat{U}_M \mid a \rangle = \sum \langle b \mid m \rangle \langle m \mid a \rangle \exp\left(-\frac{i}{\hbar} S(a,m,b)\right).
\end{equation}
Since the maximal value is obtained when all of the phases are equal, the action $S(a,m,b)$ is determined by the quantum phases of $\langle b \mid m \rangle \langle m \mid a \rangle$. To obtain a geometric phase, it is convenient to define $S(a,m,b)$ as
\begin{eqnarray}
\label{eq:SAMB}
S(a,m,b) &=& \hbar \left(\mbox{Arg}(\langle b \mid m \rangle \langle m \mid a \rangle) - \mbox{Arg}(\langle b \mid a \rangle)\right)
\\ \nonumber
&=& \hbar \mbox{Arg}(\langle b \mid m \rangle \langle m \mid a \rangle \langle a \mid b \rangle).
\end{eqnarray}
In a sufficiently large Hilbert space, $S(a,m,b)$ will be a slowly varying function of the eigenvalues $x_a$, $x_b$ and $x_m$ associated with the eigenstates $\mid a \rangle$, $\mid b \rangle$ and $\mid m \rangle$. An approximate solution of the inner product in Eq.(\ref{eq:expand}) can then be obtained by omitting all contributions with phases that oscillate rapidly in $x_m$, leaving only a small region around the least action value of $x_m$, given by
\begin{equation}
\label{eq:leastaction}
\frac{\partial}{\partial x_m} S(x_a,x_m,x_b) = 0.
\end{equation}
It is then possible to recover the classical form of causality defined by the principle of least action. Specifically, Eq.(\ref{eq:leastaction}) defines a deterministic relation between $x_m$ and $(x_a,x_b)$, so that the intermediate property can be expressed as a function $x_m(x_a,x_b)$ of the initial and final conditions.  

$S(x_a,x_m,x_b)$ can also be derived from the dynamics of states with finite uncertainties. As shown in \cite{Hib18}, the application of a unitary transformation that modifies the quantum phases by a factor of $\exp(-i x_m t/\hbar)$ moves a wave packet of energy $x_m$ from $x_a$ to $x_b$ within a propagation time of 
\begin{equation}
t(x_a,x_m,x_b) = \frac{\partial}{\partial x_m} S(x_a,x_m,x_b).
\end{equation}
Derivatives of the action $S(x_a,x_m,x_b)$ thus describe propagation times between initial and final conditions. Importantly, these propagation times constitute a macroscopic effect of the precise coherences in Hilbert space defined by Eq.(\ref{eq:SAMB}). Likewise, the principle of least action in Eq.(\ref{eq:leastaction}) is merely the macroscopic limit of the actual quantum interference effects in Eq.(\ref{eq:expand}). Specifically, the principle of least action states that the value of $x_m$ for $(x_a,x_b)$ is identified with the value of $x_m$ for which the transformation distance between $x_a$ and $x_b$ is zero. However, the vicinity of this value of $x_m$ must be included in the inner product given by Eq.(\ref{eq:expand}). Classical causality can only be recovered if $x_m$ is defined with a sufficiently low precision, so that the necessary quantum coherence in $x_m$ can be maintained. To understand this constraint, it is necessary to take a closer look at the possibility of verifying the causality relation implied by the principle of least action using an intermediate measurement.

\section{Intermediate measurements with negligible disturbance}

The main reason why the notion of a microscopic reality is so problematic in quantum mechanics is the observation that there is no non-invasive measurement by which intermediate realities could be looked up without any changes to the state of the system. In terms of the causality relation between $\mid a \rangle$ and $\mid b \rangle$, this means that any intermediate measurement relating to $\mid m \rangle$ will change the probabilities $P(b|a)$ that define the causality relation. It is therefore important to consider the conditions under which we can approximately neglect the changes to $P(b|a)$ while still obtaining meaningful information about $x_m$. 

According to the rules of quantum mechanics, an intermediate measurement of $x_m$ will change the probability amplitudes of the eigenstates $\mid m \rangle$ in accordance with the Bayesian probability update associated with the measurement outcome $r$. As discussed in \cite{Pat19}, this probability update is directly responsible for the decoherence in the system caused by the measurement interaction. At the most fundamental level, a measurement is therefore represented by the conditional probabilities $P(r|x_m)$, and the minimal decoherence is represented by an operator $\hat{M}(r)$ with
\begin{equation}
\label{eq:rb}
\langle b \mid \hat{M}(r) \mid a \rangle = \sum_m \langle b \mid m \rangle \langle m \mid a \rangle \sqrt{P(r|x_m)}. 
\end{equation}
The measurement operator $\hat{M}(r)$ extracts information about the intermediate state $\mid m \rangle$ while also modifying the causality relation between $a$ and $b$ by an unavoidable disturbance. In general, the probability of the outcome sequence $(r,b)$ is given by
\begin{equation}
P(r,b|a)=|\langle b \mid \hat{M}(r) \mid a \rangle|^2.
\end{equation}
The disturbance of the causality relation between $a$ and $b$ is negligible if the probability $P(b|a)$ is not changed by the measurement and the joint probability $P(r,b|a)$ can be written as
\begin{equation}
P(r,b|a) \approx P(r|a,b) P(b|a). 
\end{equation} 
This relation between probabilities corresponds to an analogous relation between probability amplitudes in the Hilbert space inner product of Eq.(\ref{eq:rb}). Specifically, the expectation is that the application of the operator $\hat{M}(r)$ will not change the inner product of $\mid a \rangle$ and $\mid b \rangle$, and return only one value of $\sqrt{P(r|x_m)}$ for each result $r$, 
\begin{equation}
\label{eq:approx}
\langle b \mid \hat{M}(r) \mid a \rangle \approx \sqrt{P(r|x_m)} \langle b \mid a \rangle.
\end{equation}
The most important aspect of this approximation is the selection of the value of $x_m$, which corresponds to the classical notion of an intermediate reality of $x_m$ determined by the causality relation between $a$ and $b$. According to the discussion in the previous section, most of the contributions to the sum over $m$ in Eq.(\ref{eq:rb}) cancel out because of the rapidly oscillating phases associated with the action gradient $\partial S/\partial x_m$. The multiplication with $\sqrt{P(r|x_m)}$ has no effect on this cancellation as long as $P(r|x_m)$ changes only little over one period of the phase oscillation. Likewise, the separation in Eq.(\ref{eq:approx}) is possible if $P(r|x_m)$ changes only slowly in the relevant region of nearly stationary action, where the summation over neighboring states $\mid m \rangle$ does not vanish. Eq(\ref{eq:rb}) then effectively selects the least action value $x_m(x_a,x_b)$, seemingly confirming the classical notion of causality according to which the combination of $x_a$ and $x_b$ determines the precise value of $x_m$. However, the precision of $P(r|x_m)$ is now constrained by the need to maintain quantum coherence in a sufficiently wide range of eigenstates $\mid m \rangle$. The identification of this range of eigenstates is the main purpose of the present paper. 

In order to show that the approximation in Eq.(\ref{eq:approx}) is indeed justified, we have to make use of the slow variation of $S(x_m)$, which makes it possible to approximate the sum in Eq.(\ref{eq:expand}) by an integral,
\begin{equation}
\label{eq:integral}
\langle b \mid a \rangle \approx \int \sqrt{\rho(x_m|b)\rho(x_m|a)} \exp(\frac{i}{\hbar} S(x_a,x_m,x_b)) dx_m,
\end{equation}  
where the conditional probability densities are obtained by multiplying the conditional probabilities $P(m|a)$ and $P(m|b)$ with the density of states in $\hat{M}$ given by the inverse of the eigenvalue difference $\Delta x_m$ between subsequent states $\mid m \rangle$,
\begin{equation}
\rho(x_m|\psi) = \frac{|\langle m \mid \psi \rangle|^2}{\Delta x_m},
\end{equation}
where 
\begin{equation}
\Delta x_m = x_{m+1} - x_{m}.
\end{equation}
The solution of the integral in Eq.(\ref{eq:integral}) can now be performed in the immediate vicinity of the least action solution $x_m(x_a,x_b)$. The quantum interference effects in Eq.(\ref{eq:expand}) can then be represented by 
\begin{equation}
\label{eq:separate}
\langle b \mid a \rangle \approx \frac{|\langle b \mid m\rangle \langle m\mid a \rangle|}{\Delta x_m} \int \exp\left(\frac{i}{\hbar}\left(S(x_a,x_m,x_b)+\frac{1}{2} \frac{\partial^2}{\partial x_m^2}S(x_a,x_m,x_b) (x^\prime - x_m)^2\right)\right) dx^\prime, 
\end{equation}
where $m$ and $x_m$ are the values at which the action is stationary ($\partial S/\partial x_m=0$) and the variable $x^\prime$ is used to express small variations of $x_m$ around the value at which the action is minimal. Due to the slow variation of the absolute values of $\langle b \mid m \rangle$ and of $\langle m \mid a \rangle$, the second derivative of the action $S(x_m)$ is fully determined by the inner products of the state vectors,
\begin{equation}
\label{eq:varact}
\frac{\partial^2}{\partial x_m^2}S(x_a,x_m,x_b) = \frac{2 \pi \hbar}{\Delta x_m^2} \left|\frac{\langle b \mid m \rangle\langle m \mid a \rangle}{\langle b \mid a \rangle}\right|^2.
\end{equation}
It should be noted that the Hilbert space inner products enter this relation in the form of a weak value for the projector $\mid m \rangle \langle m \mid$, highlighting the fundamental role of such weak values in defining the relations between the physical properties $x_a$, $x_b$ and $x_m$ associated with the eigenstates $\mid a \rangle$, $\mid b \rangle$ and $\mid m \rangle$ \cite{Hof12a,Hof14,Hof16}.  

If the approximation in Eq.(\ref{eq:separate}) is sufficiently accurate, it can also be applied to any intermediate measurement of $x_m$, as represented by the measurement operators $\hat{M}(r)$. The approximation given in Eq.(\ref{eq:approx}) is therefore valid whenever the conditional probabilities $P(r|x_m)$ that characterize the measurement operators $\hat{M}(r)$ vary more slowly than the phases in the integration over $x^\prime$ in Eq.(\ref{eq:varact}). This condition can be expressed in a particularly symmetric form, since both phases and probabilities are dimensionless. The separation of intermediate measurement and propagation causality expressed by Eq.(\ref{eq:approx}) is valid for
\begin{equation}
\label{eq:condition}
\frac{\partial^2}{\partial x_m^2}P(r|x_m) \ll \frac{1}{2\pi \hbar} \frac{\partial^2}{\partial x_m^2}S(x_a,x_m,x_b).
\end{equation}
We can use this condition to identify the maximal disturbance-free resolution of $x_m$,
\begin{equation}
\label{eq:resolve1}
\frac{1}{\delta x_m} = \sqrt{\frac{1}{2\pi \hbar} \frac{\partial^2}{\partial x_m^2}S(x_a,x_m,x_b)},
\end{equation}
where $\delta x_m$ is the interval around the least action value $x_m$ that contributes significantly to the inner product $\langle b \mid a \rangle$ and hence to the causality relation between $x_a$ and $x_b$. At resolutions lower than $1/\delta x_m$, the measurement results reveal the least action value $x_m$ without changing the outcome $b$ of the experiment. Within this limit, we can therefore think of $x_m$ as an intermediate reality associated with the propagation of the system from $x_a$ to $x_b$. 

Clearly, it is a necessary condition for a disturbance-free observation of the property $x_m$ that the interval $\delta x_m$ includes a large number of quantum states $\mid m \rangle$. Using Eq.(\ref{eq:varact}), we can identify the number of states in an interval of $\delta x_m$ and determine the limit of quantum state resolution,
\begin{equation}
\label{eq:resolve2}
\frac{1}{\delta n} = \frac{\Delta x_m}{\delta x_m} = \left|\frac{\langle b \mid m \rangle\langle m \mid a \rangle}{\langle b \mid a \rangle}\right|.
\end{equation}
Classical intermediate realities therefore emerge only if the inner products between the different eigenstates are sufficiently low. This observation highlights a fundamental problem of quantum information theory: the focus on individual states and low dimensional Hilbert spaces makes it impossible to relate the results obtained in this extreme limit of quantum mechanics to the more familiar physics of cause and effect that governs the technology used to control the quantum system. It thus remains a challenging task to properly explain the fundamental nature of causality in terms of quantum interference effects, without any redundant references to intermediate realities. The analogy between eigenstates and classical information that is widely used to present quantum information technologies as the next generation of conventional computers seems to be rather misguided in that respect.   

\section{Failures of the principle of least action}

The results presented above explain why it is sometimes possible to describe a sequence of observations by unitary transformations even though the intermediate measurements must be represented by self-adjoint operators $\hat{M}(r)$ that change the quantum state in a non-unitary manner. It has been shown that the intermediate observations will then be consistent with the intermediate values $x_m$ for which the action $S(x_a,x_m,x_b)$ is minimal. It should be noted that the action $S(x_a,x_m,x_b)$ is consistent with the Lagrangian action for minimal action paths from $x_a$ to $x_m$ and from $x_m$ to $x_b$ \cite{Hof12a,Har09,Dre14b}. In some sense, the present argument is therefore a more compact formulation of the relation between quantum phases and the action known from the path integral formalism. However, it should be noted that the Lagragian approach vastly over determines the time evolution by unnecessarily assigning values to an infinite number of unobserved quantities at the same time \cite{Hof12a}. The present analysis shows why such an assignment is redundant by examining its empirical justification for the more reasonable case of a single unobserved quantity $x_m$. The breakdown of the principle of least action can then be identified and characterized in terms of the modification of causality from the approximate assumption of intermediate realities to the more accurate understanding of a causality relation mediated by quantum coherences. The essential effect concerns the rules that govern the modifications of $P(b|a)$ caused by more precise observations of $x_m$. These rules show that the disturbance associated with the measurement process is determined by the action $S(x_a,x_m,x_b)$ and is therefore an intrinsic feature of the objective causality governing the undisturbed motion. I would therefore argue that the causality relation between different observations is completely objective, even when an interaction relating to one of the observables changes the causality relations between the other two. At the same time, the concept of an internal reality of the observables loses its meaning, since the objective nature of the disturbance means that the observation of $x_m$ changes the role of $x_m$ in the causality relation between the initial observation $x_a$ and the final relation $x_b$. It seems to be an exercise in futility to attempt a separation between the reality of $x_m$ as an external effect and the reality of $x_m$ as a modification in the causality relation between $x_a$ and $x_b$, since quantum mechanics provides a perfectly reasonable explanation of their effective entanglement (see also \cite{Pat19}). 

To make the argument more specific, it might help to analyze Eq.(\ref{eq:rb}) in the limit of high resolution. In this limit, Eq.(\ref{eq:approx}) loses its validity and the principle of least action is violated. Instead, the complete action function $S(x_a,x_m,x_b)$ contributes to the probability of the outcome $\mid b \rangle$ in a manner that is fully determined by the resolution of the intermediate measurement. Specifically, Eq.(\ref{eq:rb}) can be solved by an integral localized in a small region around the measurement outcome $x_r$ that describes the maximal Bayesian likelihood obtained from the conditional probability $P(r|x_m)$. The condition that the measurement resolution $1/\delta x_r$ is higher than the resolution limit of $1/\delta x_m$ for non-disturbing measurements means that the second derivative in the action can be neglected, so the integral can be written as
\begin{equation}
\label{eq:highres}
\langle b \mid \hat{M}(r) \mid a \rangle \approx \frac{\langle b \mid m \rangle \langle m \mid a \rangle}{\Delta x_m} \int \sqrt{P(r|x^\prime)} \exp\left(\frac{i}{\hbar} \frac{\partial}{\partial x_m} S(x_a,x_m,x_b) (x^\prime-x_r)\right)dx^\prime.
\end{equation}
Essentially, the integral corresponds to a Fourier transform of the resolution function $\sqrt{P(r|x_m)}$, where the Fourier component is determined by the gradient of the action at $x_m=x_r$. What is being resolved in the measurement is not the value of $x_m$, but the action gradient associated with the external effect $x_r$. In the case of intermediate measurements with Gaussian resolution, the approximate result for the measurement sequence is
\begin{equation}
\label{eq:Sgrad}
\langle b \mid \hat{M}(r) \mid a \rangle \approx \langle b \mid m \rangle \langle m \mid a \rangle \left(8 \pi \frac{\delta x_r^2}{\Delta x_m^2} \right)^{1/4} \exp\left(- \left(\frac{\delta x_r}{\hbar} \frac{\partial}{\partial x_m} S(x_a,x_m,x_b)\right)^2\right).
\end{equation}
If additional sources of decoherence are avoided, the measurement outcomes $x_r$ can provide rather detailed information on the action gradients that govern the causality relation between $x_a$ and $x_b$ in Hilbert space. The failure to observe the intermediate realities $x_r=x_m$ of the undisturbed propagation from $x_a$ to $x_b$ therefore originates from the role that small action gradients play at the microscopic level. As shown by Eq.(\ref{eq:Sgrad}), the relevant condition for the failure of the least action approximation is
\begin{equation}
\frac{1}{\delta x_r} > \frac{1}{\hbar} \left|\frac{\partial}{\partial x_m} S(x_a,x_m,x_b) \right|. 
\end{equation}
Quantum theory thus applies in the limit of resolutions larger than the action gradient evaluated in units of $\hbar$. This statement is as fundamental to quantum theory as the statement that the theory of relativity applies at velocities approaching the speed of light is fundamental to the theory of relativity. It clearly identifies the magnitude of the effects described by the theory and therefore explains why they can be neglected in the classical limit. I would therefore conclude that quantum mechanics describes the details of causality relations in the limit of high resolution, where the action provides a universal measure of resolution for causality relations in all fields of physics. The main problem that has prevented us from understanding quantum physics as the natural foundation of classical physics is that we are not used to a quantification of causality in terms of the action. It is therefore necessary to carefully consider the role of the action in unitary dynamics and its relation with the concept of quantum coherence as shown in Eqs.(\ref{eq:unitary}) and (\ref{eq:SAMB}). Ultimately, the role of the action as a universal expression of causality is the cornerstone of a proper understanding of the physical world at the microscopic limit.  

\section{Conclusions}

As I have explained above, it is possible to understand the physics described by Hilbert space inner products as a universal description of causality at the ultimate limit of quantitative precision. The action emerges naturally from the complex phases that appear in Hilbert space products when they are expressed in terms of components that seem to represent possible intermediate realities. However, these intermediate contributions cannot be converted into observable realities without changing the original causality relation between initial conditions and final measurement. Instead, the necessary modification of causality relations caused by any intermediate observation is fully determined by the action $S(x_a,x_m,x_b)$, which provides a complete description of deterministic causality relations between the different physical properties of a system \cite{Hof12a}. By focusing the discussion on causality relations between physical properties that cannot be measured jointly, it is possible to make statements about causality without any speculation about the nature of quantum states. The result is a theory that manages to smoothly connect the approximate description of phenomena by classical physics to the more precise description provided by quantum mechanics without changing the conceptual framework. I would argue that this result can reconcile our classical intuition with the weirder aspects of quantum physics in a constructive manner and provide a consistent description of both quantum mechanics and its classical limit \cite{Hof15}. In particular, it should not come as such a great shock that the naive assumption of a microscopic reality breaks down as a consequence of the fundamental action scale given by $\hbar$. Even in the classical limit, we merely reconstruct the reality ``out there'' from observations that are never very precise. The possibility to do so depends entirely on the reliability of causality relations such as the one represented by the Hilbert space inner product in Eqs.(\ref{eq:expand}) and (\ref{eq:rb}). What I have shown here is that the classical versions of causality are robust up to a resolution of $\delta x_m$ given by the curvature of the action at its minimum. It may be worth noting that even a measurement with a resolution much lower than $1/\delta x_m$ modifies the quantum state $\mid a \rangle$ significantly by removing all amplitudes $\langle m \mid a \rangle$ outside of the interval $\delta x_m$. The criterion for classical causality is therefore less restrictive than the criterion for quantum state disturbance, and this fact explains why many forms of quantum coherence have no observable effects whatsoever. For instance, the quantum coherence of a superposition of dead cats and living cats famously suggested by Schr\"odinger has no observable consequences in the future, and this will even be true for the experience of the cat itself. Quantum corrections of classical causality relations can only be observed if both state preparation and measurement are sufficiently precise, since neither one has any physical meaning of its own. The quantitative nature of quantum corrections of causality can then be observed and quantified in terms of the statistical relations between different measurement outcomes, an example of which has been given in \cite{Hof17,Hof18,Hof19} for the failure of Newton's first law in the case of particle propagation in free space. A closer inspection of the relation between causality and quantum coherence can thus result in the systematic development of new means of control beyond the least action approximation. 

In the light of the results presented above, it seems that the idea that quantum states and their eigenvalues can describe the physical reality of an object is based on the misconception that the reality of an object can be separated from the causality relations that are necessary to experience that reality. The answer to all interpretational problems of quantum mechanics should therefore lie in an improved understanding of the causality relation between objects and their observable effects, where the action can take its rightful place as a fundamental scale in all physical theories. The analysis given above evaluates the precise quantitative limits for the emergence of a classical reality in quantum causality relations. I hope that the discussion presented in this paper will thus prove to be the first step towards a deeper understanding of quantum theory as the most fundamental explanation of all observable phenomena. 

\section*{Acknowledgment}
This work has been supported by JST-CREST (JPMJCR1674), Japan Science and Technology Agency.

\end{document}